\documentclass[twocolumn,showpacs,preprintnumbers,a4paper,prl,floatfix]{revtex4}

\usepackage{graphicx}%
\usepackage{dcolumn}
\usepackage{amsmath}
\usepackage{amssymb}
\usepackage{bm}

\makeatletter
\def\btt#1{\texttt{\@backslashchar#1}}%
\DeclareRobustCommand\bblash{\btt{\@backslashchar}}%
\makeatother

\begin{document}

\preprint{ca.tex}

\title{Influence of carbon doping in the vortex matter properties of MgB$_2$}

\author{M. Pissas, D. Stamopoulos}
\affiliation{Institute of Materials Science, NCSR,  Demokritos,
15310 Aghia Paraskevi, Athens, Greece}

\author{S. Lee, S. Tajima}
\affiliation{Superconductivity Research Laboratory, ISTEC, Tokyo 135-0062, Japan}%

\date{\today}
\begin{abstract}

We have studied the vortex-matter phase diagram of low C doped
MgB$_2$ superconductor. The significant finding is the appearance
of a peak effect, which, as both our global magnetization and
local ac susceptibility measurements revealed, is placed well
below the $H_{c2}$-line. The absence of significant bulk pinning
below the onset line $H_{on}(T)$ of the peak effect implies that
the Bragg glass phase is present for $H < H_{on}(T)$.
Interestingly, the unexpected absence of bulk pinning above the
end point line $H_{ep}(T)$ of the peak effect implies the presence
of a slightly pinned vortex phase in the regime $H_{ep}(T) < H <
H_{c2}(T)$. In addition, the observed increase of the
$H_{c2}^c(0)$ since the carbon doped MgB$_2$ becomes more dirty
and the reduced anisotropy, in comparison to the pure MgB$_2$,
makes the C-doped MgB$_2$ more favorable for practical
applications.

\end{abstract}
 \maketitle

In spite of the rich physics of recently discovered MgB$_2$
\cite{nagamatsu01}, due to the multi component order parameter
\cite{twogap}, the potential of MgB$_2$ for applications is
limited by comparatively low upper-critical fields
$H_{c2}^c(0)\approx 30$ kOe \cite{pissas02}. Atomic substitutions
may influence the basic properties of a superconductor making it
appropriate for practical applications, by increasing the
$H_{c2}(T)$ line and/or the critical current that can sustain
without losses. On the other hand, atomic substitutions may help
in the clarification of a number of issues related with the basic
mechanism which is responsible for the superconductivity in the
particular compound. It has been theoretically proposed that the
$H_{c2}(T)$ line can be increased by adding nonmagnetic impurities
\cite{gurevich03}. The MgB$_2$ crystal structure consists of
alternating close packed Mg$^{+2}$ layers and honeycomb-like boron
sheets. $T_c$ was found to gradually decrease upon substitution of
Mg for Al and B for C, consistent with a decrease in the
density-of-states at the Fermi level induced by electron doping
and reduced lattice volume. Although both Al and C reduce the
$T_c$, the filling of the $\sigma$-band may reduce the anisotropy
of the pristine MgB$_2$,\cite{anisotropy} helping  this way the
tuning of the basic properties of MgB$_2$.

Single crystals were grown by a high-pressure technique
\cite{lee03} previously developed for the growth of pristine
MgB$_2$ phase \cite{lee01} using precursors with a nominal
composition of Mg(B$_{1-x}$C$_{x}$)$_2$ $x=0.02-0.20$. The
residual resistivity ratio at  $T_c$, of crystals from the same
batch, as  used in the present work is
$\rho_{ab}(300)/\rho_{ab}(T_c)\approx 2.2$ while the residual
resistivity increases by a factor of 3-4 in comparison to the
pristine MgB$_2$, indicating a degreasing of the mean free path by
a factor of 3-4.

Local ac susceptibility measurements were carried out on a
MgB$_{1.96}$C$_{0.04}$ single crystal by means of a GaAsIn Hall
chip (active area of $50\times 50$ $\mu $m$^2$). The dimensions of
the crystal are $750\times 350\times 40$ $\mu$m$^3$ with the
shorter length along the $c$-axis. The crystals have $T_c$ of 35.6
K at zero dc field ($H_{ac}=1 $ Oe) and a transition width of 0.3
K (10\%-90\% criterion). In order the small ac field to be
measured, under the presence of a large dc magnetic field, a
second sensor of the same size was connected opposite to the first
one by means of an ac bridge. The real and imaginary parts
($V=V'+iV''$) of the modulated Hall voltage, which is proportional
to the local magnetic moment ($V\propto m\propto B-H_{\rm ac}$) at
the surface of the crystal, were measured by means of two lockin
amplifiers.
DC magnetization measurements were performed (SQUID) magnetometer.

\begin{figure}[htbp]\centering
\includegraphics[angle=0,width=\columnwidth]{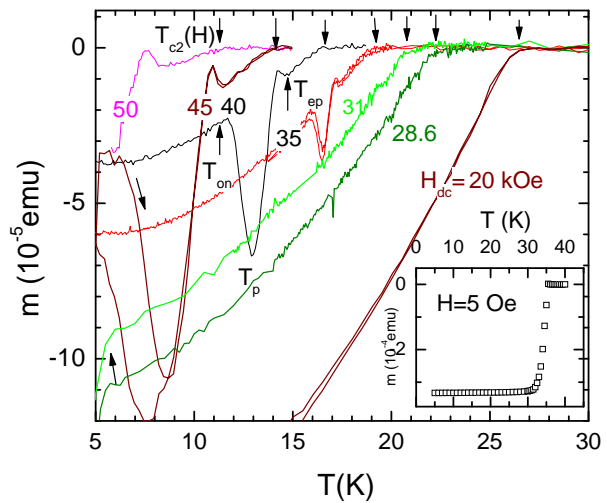}
\caption{Zero field and field cooled temperature variation of the
global magnetic moment of MgB$_{1.96}$C$_{0.04}$  crystal for
${\bf H}\parallel c$-axis. The inset shows the temperature
variation of the magnetic moment for $H=5$ Oe. The small positive
peak at the region of $T_{ep}$ is an artifact due to field
inhomogeneity.} \label{fig1}
\end{figure}

Figure \ref{fig1} shows the temperature variation of the bulk
magnetic moment of the MgB$_{1.96}$C$_{0.04}$ for several magnetic
fields (${\bf H}\parallel c$) as it was measured by the SQUID
magnetometer. For small fields (1-32 kOe) the $m(T)$ curves
doesn't show any special feature and are terminated at the $m=0$
axis in a parallel fashion. The diamagnetic onset temperature is
identified with $T_{c2}(H)$ that is determined by extrapolating
the low temperature curve to $m=0$. As one can see the
irreversibility is negligible, and only when the temperature is
reduced enough, irreversible behavior is appeared. Measurements in
higher fields ($H>32$ kOe) clearly show a negative peak whose its
height and width increase as the magnetic field increases. We
define the temperatures where the peak effect starts as
$T_{on}(T)$, takes its minimum value as $T_p(T)$ and finally ends
as $T_{ep}(H)$. It is interesting to note that in the interval
$T_p<T<T_{c2}(H)$, the magnetization is nearly reversible (see the
curve measured at $H=35$ kOe) while for $T_{on}(H)<T<T_p(H)$ the
magnetic moment presents hysteretic behavior which increases as
the magnetic field increases.

For a type-II superconductor, near $H_{c2}$ line is expected that
the magnetization varies as $4\pi
M=-(H_{c2}(T)-H)/(2\kappa_2^2-1)\beta_{\rm A}\approx
(H_{c2}(T)-H)/(2\kappa_2^2\beta_{\rm A})$ \cite{abrikosov}, where
$\beta_{A}$ is determined by the geometrical arrangement of
fluxoids in the mixed state and $\kappa_2(T)$ is the second
Ginzburg-Landau-Maki parameter\cite{maki}. The isofield $M(T)$
slope, near $T_{c2}$ is determined by two factors: $\partial
M/\partial T|_{H_{dc}}=-(1/2\kappa_2^2(T)\beta_A)[dH_{c2}/dT-
+(1/\kappa_2)(d\kappa_2(T)/dT)(H_{c2}(T)-H)]$. Since
$\kappa_2\approx 20-30$ one can ignore the second term,
consequently, the slope of the magnetization curves near $T_{c2}$
is mainly determined from the $H_{c2}(T)$ line slope. The observed
reduction of the magnetization slope as the field increases means
that the $H_{c2}(T)$ line goes to $T=0$ with decreasing slope.

\begin{figure}[htbp]\centering
\includegraphics[angle=0,width=\columnwidth]{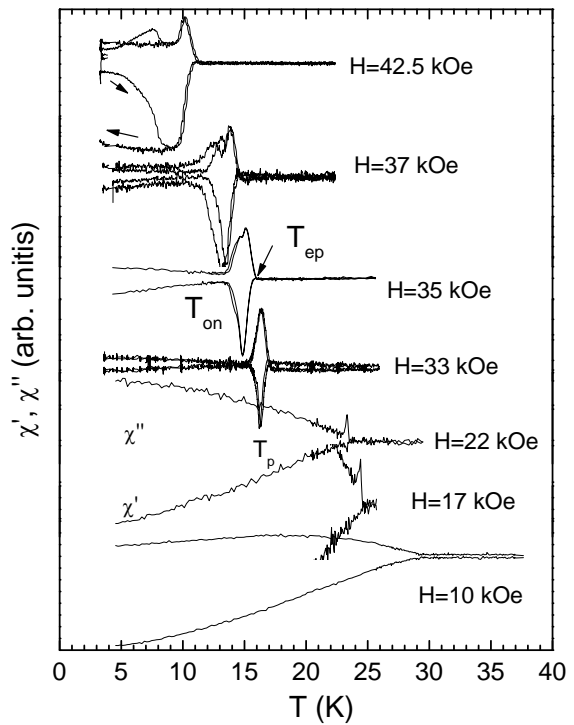}
\caption{ Real ($\chi'$) and imaginary ($\chi''$) local
fundamental ac-susceptibility as a function of temperature, for
$H_{ac}=8.8$ Oe and ($H_{dc}=10-42.5$ kOe). For $H>20$ kOe the
peak effect is always observed.} \label{fig2}
\end{figure}

Fig. \ref{fig2} shows the real and imaginary parts of the local
fundamental ac-susceptibility  ($\chi'=(B'-H_{ac})/H_{ac}$,
$\chi''=(B''-H_{ac})/H_{ac}$), as function of temperature,
measured under several dc magnetic fields $10\leq H_{dc}\leq 42.5$
kOe for an ac-field $H_{ac}=8.8$ Oe. The measurements have been
taken during cooling and heating. As the measurements are made in
higher magnetic fields both $\chi'$ and $\chi''$ form a peak.
Moreover, the location of the peak temperature $T_p(H)$ and the
end point $T_{ep}(H)$ do not depend on the amplitude of the
ac-field. Our local Hall measurements enabled us to detect the
peak effect in much lower fields than the global SQUID
measurements. In addition, below a characteristic point $(T_1,
H_1)\approx $ (24.5 K, 20 kOe) the peak effect could not be
observed. Below this characteristic point the peak effect turned
into a sharp drop which we define it by its onset $T_{on}$ and its
end point $T_{ep}$. This feature also could not be observed below
a second characteristic point $(T_2,H_2)\approx$(26 K,12 kOe) (see
Fig. \ref{fig4} below).
\begin{figure}[htbp]\centering
\includegraphics[angle=0,width=\columnwidth]{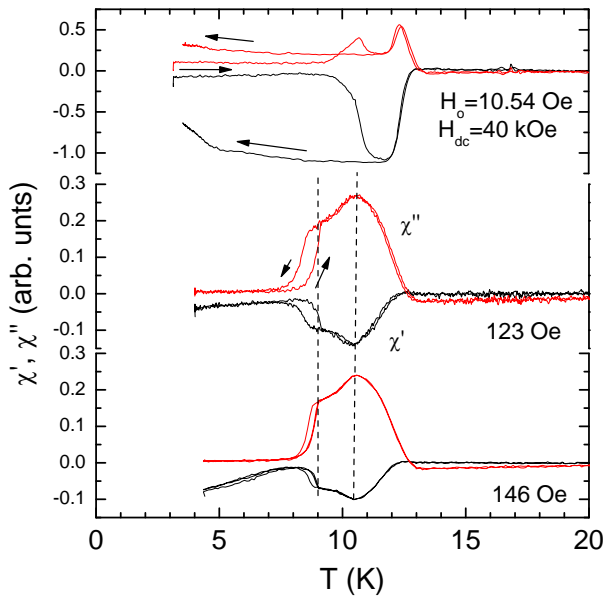}
\caption{Dependance of the hysteretic local ac-susceptibility for
$H_{dc}=40$ kOe and for various amplitudes of the ac-field.}
\label{fig3}
\end{figure}
Comparing the temperature where the diamagnetic signal appears in
our local ac susceptibility measurements to the corresponding bulk
dc measurements we find out that this doesn't correspond to the
$T_{c2}$. Here, we would like to emphasize that for the case of a
type-II superconductor in the mixed state with negligible pinning,
as the $H_{c2}$ line is approached one expects that the local
ac-susceptibility  is mainly in-phase to the ac field (for
sufficiently low frequencies) and positive (paramagnetic)
$\chi'=4\pi(dM/dH)_{H_{\rm dc}}\propto
1/2\kappa_2^2(T)\beta_{a}>0$.
We did not observe paramagnetic ac moment for $T>T_{ep}$ probably
due to the fact that, either the paramagnetic moment is below our
sensitivity limit, or we have a superposition of a paramagnetic
and a diamagnetic ac moment (due to a small critical current) of
equal size giving a zero net ac-moment.

As we presented in Fig. \ref{fig2} pronounced hysteretic behavior
of the local ac susceptibility is observed in the regime below the
peak points, when high dc magnetic fields are applied (e.g $H>35$
kOe). The appearance of hysteresis in non-equilibrium states may
be indicative of a first-order phase transition but doesn't
necessarily provide a conclusive evidence. However, while the
exact mechanism of the peak effect is not fully understood, recent
experiments on Nb\cite{ling01},
Bi$_2$Sr$_2$CaCu$_2$O$_{8+\delta}$,\cite{Avraham}
HgBa$_2$CuO$_4$,\cite{Stamopoulos02A,Stamopoulos02B}, NbSe$_2$
\cite{Marchevsky} and (K,Ba)BiO$_3$, \cite{klein01} have
correlated the onset of the peak effect with a first-order
order-disorder transition of the vortex lattice.
A high enough driving force exerted on vortices may uncover
different behavior than the one observed in a low applied
perturbation. We systematically investigated the influence of the
amplitude of the applied ac field on the hysteretic response. In
Fig. \ref{fig3} we present the temperature variation of the real
and imaginary part of local ac-susceptibility under 40 kOe
dc-magnetic field, for appreciate large amplitudes of the
ac-field. We see that the detected hysteresis progressively
reduces as we apply higher ac fields. For high enough ac fields
the hysteresis is confined only in the regime between the onset of
the peak and a new characteristic point, at which the in-phase
signal presents a kink. These results may be explained as
following: The fact that the zero field cooled vortex state
exhibits zero screening current in the regime $T<T_{on}$ suggests
that no metastable states are present in this regime and that the
so-called {\it Bragg glass state \cite{bragg} is a true
equilibrium phase.} Furthermore, we performed relaxation and
partial loop measurements in the regime between the onset point
and the peak effect. Those measurements, which will be presented
elsewhere\cite{PissasC}, revealed that partial sub-loops are
always present and that the field cooled vortex state exhibits
strong relaxation, while the zero field cooled one doesn't relax
even in long experimental times. These results indicate that the
zero field cooled vortex state is an equilibrium state while the
field cooled phase is a supercooled metastable one. In addition,
these {\it two vicinal vortex states (namely the disordered and
the Bragg glass) coexist in the finite temperature interval
$T_{on}<T<T_p$ around the transition regime}. All the above
mentioned results indicate that the detected transition is of
first order. Recently, the same behavior was detected in
YBa$_2$Cu$_3$O$_7$,\cite{Zhukov}
Bi$_2$Sr$_2$CaCu$_2$O$_{8+\delta}$ \cite{Beek00,Avraham},
HgBa$_2$CuO$_{4+\delta}$ \cite{Stamopoulos02A,Stamopoulos02B},
NbSe$_2$ \cite{Marchevsky} and V$_3$Si,\cite{gapud03} single
crystals.
\begin{figure}[htbp]\centering
\includegraphics[angle=0,width=\columnwidth]{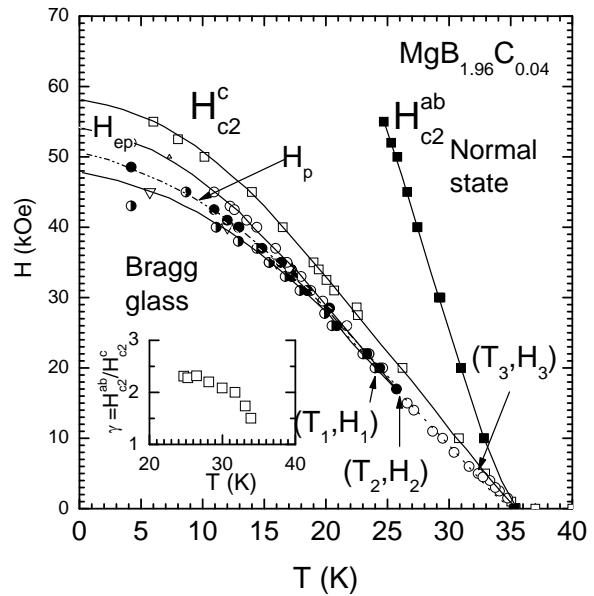}
\caption{The phase diagram of vortex matter of
MgB$_{1.96}$C$_{0.04}$. Squares and triangles come from global
SQUID measurements while circles come from local Hall
measurements. Presented are the onset line $T_{on}$ (semi-filled
circles), the peak effect line $T_p$ (solid circles), the end
point line $T_{ep}$ (open circles) and the $H_{c2}^c(T)$ (open
squares), $H_{c2}^{ab}(T)$ (solid squares) lines. The inset shows
the temperature variation of the $\gamma=H_{c2}^{ab}/H_{c2}^c$.}
\label{fig4}
\end{figure}

Figure \ref{fig4} summarizes our results in the form of $T-H$
phase diagram of the mixed state of MgB$_{1.96}$C$_{0.04}$.
A similar phase-diagram has been also determined for the $x=0.05$
sample. The main differences with $x=0.02$ sample are the larger
characteristic fields $H_{c2}^{ a,ab}$ , $H_{\rm on}, H_{\rm ep}$,
and  $H_{p}$. Remarkably, the $H_1$ ($\sim 15$ kOe) field does not
change between the pristine and carbon doped crystals.
For completeness
the $H_{c2}^{ab}(T)$-line  was added, permitting the estimation of
the $H_{c2}$ anisotropy which is depicted in the inset. This part
of the anisotropy curve is well below the corresponding curves
reported in the literature\cite{anisotropy} for the pristine
MgB$_2$\cite{c-anisotropy}. In the results presented in Fig.
\ref{fig4} there is a remarkable experimental finding. The end
point line as defined from the end point of the peak effect
doesn't coincide to the upper-critical field line as that defined
from the bulk magnetization measurements. In addition, the portion
of the mixed state occupied by the peak effect shrinks and the
magnitude of the critical current is reduced as the field
decreases. The peak effect disappears at $(T_1,H_1)$ and is
transformed to a very narrow diamagnetic step, up to the point
$(T_2,H_2)$. As the field is further reduced, the local $\chi'$
shows a monotonic conventional behavior. However, the line defined
by the diamagnetic onset points of the $\chi'$ curves is also
placed below the $H_{c2}^c$ line and only at a particular point
$(T_3,H_3)$ changes slope and coincides to the $H_{c2}^c$ line.
Based on our results and with what already is known in other
compounds the onset peak-effect line concerns an order-disorder
transition, most probably of first-order\cite{Avraham,Marchevsky},
so the $(T_2,H_2)$ is a tricritical point where a first order
transition is terminated. One could claim that $(T_2,H_2)$ point
is a critical point. However a critical point can exist only for
phases such that the difference between them is purely {\it
quantitative}. Therefore, these vortex solid phases cannot be
continuously transformed into each other and a termination of the
phase transition line, separating these phases, can terminate only
on another phase transition line\cite{mikitik01,giamarchi95}.
This fact may not be valid in our case, since the Bragg glass
phase is also {\it qualitatively} different from the phase that
occupies the region between the $H_{c2}^c$ and the end-point line.
This region of the phase diagram may be related with a amorphous
vortex phase.
We would like to stress that in this regime the critical current
is very low (normally one expects that the critical current
reduces smoothly towards the zero values at $H_{c2}^c$) and is an
important new experimental finding which needs a theoretical
explanation.
A critical point therefore cannot exist for such phases, and the
equilibrium curve must either go to $T_c(H=0)$ or terminate by
intersecting the equilibrium curves of other phases. Consequently,
the onset line below $(T_2,H_2)$ may concern a curve of
second-order phase transitions in the $T-H$ plane that separates
phases of different symmetry, namely a transition line from a
Bragg glass to an amorphous vortex phase. Alternatively, if one
supposes that the diamagnetic onset of the local susceptibility
(for $H<H_2$, see Fig. \ref{fig4}) does not represent a phase
transition but simply marks an irreversibility line, then the
point $(T_2,H_2)$ is a critical point with the consequence of a
reentrance of the Bragg glass phase below the $(T_2,H_2)$ point in
the regime $H_{ep}<H<H_{c2}^c$. At the end, we would like to
comment on the presence of the characteristic point
$(T_3,H_3)\approx$(33 K, 5 kOe). As we observe in Fig. \ref{fig4}
for magnetic fields below $H_3$ the onset of the diamagnetic local
ac response coincides to the onset of the diamagnetic global dc
magnetic moment. Although tunnelling data are not available at the
presence, we know that in pristine MgB$_2$ near $T_c(H=0)$ the
$\pi$ gap closes at about 4-5 kOe. Therefore, it is interesting to
correlate this point with the closing of the $\pi$ gap.

In summary, we experimentally estimated the vortex matter phase
diagram  for MgB$_{1.96}$C$_{0.04}$ single crystal superconductor
which exhibits half the anisotropy of the pristine MgB$_{2}$ one.
The peak effect is observed in our local Hall and in bulk SQUID
measurements. The respective line $H_p(T)$ and even the end point
line $H_{ep}(T)$ are placed well below the upper-critical field
line $H_{c2}(T)$. The $H_p(T)$ line doesn't terminate on the
$H_{c2}(T)$ line but disappeared at $(T_2,H_2)$. The peak effect
line concerns an order-disorder transition of first-order. This
line probably is terminated at the tricritical point $(T_2,H_2)$
continuing as a line of second order transitions up to $T_c$. The
$H_{c2}^c(0)$ is obviously increased in the C doped MgB$_{2}$
crystal, implying a decreased mean free path of the charge
carrier.

\begin{acknowledgments}
This work was partially supported by the New Energy and Industrial Technology Development
Organization (NEDO) as collaborative research and development of fundamental technologies
for superconductivity applications.
\end{acknowledgments}

\end{document}